\documentclass[notitlepage,nofootinbib,preprintnumbers,amssymb,superscriptaddress]{revtex4-1}
\usepackage{amssymb,float,amsmath,amsfonts,amsthm,cases,bm,latexsym,mathrsfs, eufrak, graphicx, color, epsfig, wrapfig, comment, xcolor, titlesec}
\usepackage[normalem]{ulem}
\usepackage{cancel}
\definecolor{linkcolor}{HTML}{799B03}
\definecolor{urlcolor}{HTML}{799B03}
\definecolor{ultramarine}{rgb}{0.07, 0.04, 0.56}
\definecolor{cadmiumgreen}{rgb}{0.0, 0.42, 0.24}
\definecolor{indigo(dye)}{rgb}{0.0, 0.25, 0.42}
\usepackage[linktocpage=true]{hyperref}
\hypersetup{
colorlinks=true,
citecolor=green,
linkcolor=cadmiumgreen,
urlcolor=indigo(dye),
}

\usepackage{autobreak}

\def\[{\begin{equation}}
\def\]{\end{equation}}

\def\d{\partial}

\begin{document}

\title{No-Go theorem in the cubic subclass of Horndeski theory for spherically symmetric dynamical background.}

\author{S. Mironov}
\email{sa.mironov\_1@physics.msu.ru}
\affiliation{Institute for Nuclear Research of the Russian Academy of Sciences,
60th October Anniversary Prospect, 7a, 117312 Moscow, Russia}
\affiliation{Institute for Theoretical and Mathematical Physics,
MSU, 119991 Moscow, Russia}
\affiliation{NRC, "Kurchatov Institute", 123182, Moscow, Russia}

\author{M. Sharov}
\email{sharov.mr22@physics.msu.ru}
\affiliation{Institute for Nuclear Research of the Russian Academy of Sciences,
60th October Anniversary Prospect, 7a, 117312 Moscow, Russia}
\affiliation{Institute for Theoretical and Mathematical Physics,
MSU, 119991 Moscow, Russia}

\begin{abstract}
We consider a general dynamical, spherically symmetric background in the cubic subclass of Horndeski theory and obtain the quadratic action for the perturbations using the DPSV approach. We analyze the stability conditions for high-energy modes and study the issue of the no-go theorem. We formulate the no-go theorem for weak dependence on one variable (time or radial) and derive its generalization to the cases which could be reduced by coordinate transformation to scenarios where the scalar field has weak dependence on one of the coordinates. Moreover we show that wide class of singular solutions is also prohibited within the cubic subclass of Horndeski theory.
\end{abstract}

\maketitle
\section{Introduction}
\hspace{\parindent} Scalar-tensor theories (including Horndeski theory) represent a group of models of modified gravity and have a wide range of applications. We study the stability of classical solutions in Horndeski theory, which is the most general scalar-tensor theory of gravity with an additional scalar field and second-order equations of motion, which in turn guarantees the absence of Ostrogradski ghosts \cite{Horndeski, Trincherini:2008, Deser:2009, Covariant_Galileon, Koba_G-infl}, see \cite{Koba_review} for a review. This class of theories is interesting due to the possibility of violating the Null Energy Condition (NEC) \cite{NEC}, which subsequently allows the existence of stable (at least locally) solutions with modified gravity such as various cosmologies without singularities \cite{Creminelli:2006, DPSV, G-bounce2011, Creminelli:2010, Creminelli:2012, Cai:2012} and compact objects including traversable wormholes \cite{Rubakov:2015, 1601R}. To build any physically acceptable solution, it is necessary to check the stability conditions, and this is the main subject of our work. 

So far, a full stability analysis in Horndeski theory, at least for the high-momentum regime, has been performed only for static spherically symmetric and cosmological backgrounds \cite{Kobayashi:odd, Kobayashi:even, Koba_G-infl, Arina_ani}. The structure of the stability conditions in these cases allows one to formulate the no-go theorem for a fully stable non-singular solution for a static spherically symmetric background \cite{1601R, Olegi} and cosmological background \cite{Cosmologies:2016, Kobayashi_no-go, math_nogo} in Horndeski theory, see \cite{review_nogo} for a review of studies on healthy solutions in scalar-tensor theories. The no-go theorem was also proved within the context of multi-Galileon theory \cite{Kolevatov_2_scalar, Koba_multi_gal}. Existing studies partly cover the case of a general dynamical spherically symmetric background in which perturbations were considered only in the odd-parity sector (according to the Regge-Wheeler classification of perturbations \cite{Regge}) of Horndeski theory \cite{odd_dynamic, Kaluza_odd}. The case of a shift-symmetric scalar field (i.e. $\pi(r,t) = q \cdot t + \psi(r))$ with static background metric functions was separately considered due to its connection to hairy black hole solutions \cite{Takahashi:2016dnv, Babichev_sol}. However, the stability analysis in the just outlined case was performed only for the odd-parity sector \cite{Ogawa}.

We consider a general dynamical, spherically symmetric background in the subclass $L_2 + L_3$ (cubic subclass) of Horndeski theory \cite{1601R, G-bounce2011}. The only perturbation in the cubic subclass of Horndeski theory that differs from GR is the scalar mode, which contributes to the even sector of perturbations, both tensor modes are similar to GR because the gravity is not modified in this subclass. For simplicity, we analyze the behavior of linear perturbations using the DPSV approach \cite{DPSV, DPSV:1708}. In higher subclasses of Horndeski theory, the DPSV trick breaks down for a spherically symmetric background \cite{DPSV_2023, DPSV:1712} and it is not usable to derive quadratic action in full Horndeski theory.

The background we consider here is covered by the ADM formalism and the method of Hamiltonian analysis \cite{ADM_Langlois}, see \cite{Langlois_review} for a review, but we do not adopt it and work in covariant formalism. Note that the unitary gauge which is often imposed with the ADM formalism assumes that the scalar field gradient is timelike at any point. In this paper, we do not make such an assumption.

The Lagrangian of the cubic subclass of Horndeski theory reads as
\begin{equation} \label{Lagrangians}
    L =   - \frac{1}{2k}R + F(\pi, X) + K (\pi, X) \Box \pi  \; ,
\end{equation}
where $k = 8 \pi G$, $R$ is the Ricci scalar, $\pi$ is the Galileon field, $F$ and $K$ are arbitrary Lagrangian functions, and $X = \nabla_\mu \pi \nabla^\mu \pi$, $\Box \pi =  \nabla_\mu  \nabla^\mu \pi$.

\section{Stability conditions}
\hspace{\parindent} In the cubic subclass of Horndeski theory, it is possible to integrate out metric perturbations from the quadratic action using the Einstein equations \cite{DPSV, 1601R}. This method (DPSV approach) allows one to calculate the quadratic action for the second derivatives of the perturbation $\chi$ of the scalar field $\pi$, and it can be used to obtain high-momentum stability conditions, which we derive in this section. 

The Galileon energy-momentum tensor reads

\begin{equation} \label{emt}
T_{\mu \nu} = 2 F_X \d_\mu \pi \d_\nu \pi +  2 K_X \Box \pi \cdot \d_\mu \pi \d_\nu \pi - \d_\mu K \d_\nu \pi - \d_\nu K \d_\mu \pi - g_{\mu \nu} F + g_{\mu \nu} g^{\lambda \rho} \d_\lambda K \d_\rho \pi \; ,
\end{equation}

where $F_\pi = \d F/\d \pi$, $F_X = \d F/\d X$,
etc.,
and $\d_\mu K = K_\pi \d_\mu \pi + 2 K_X \nabla^\lambda \pi \nabla_\mu
\nabla_\lambda \pi$.

To derive the quadratic Lagrangian for the Galileon perturbations of
high momenta and frequencies we write the linearized Galileon equation \cite{DPSV} with neglected terms without second derivatives
\begin{align}
 -2 [F_X  + K_X \Box \pi - K_\pi +  \nabla_\nu (K_X \nabla^\nu \pi)]
 \nabla_\mu \nabla^\mu \chi  &
\nonumber \\
-2  [2 (F_{XX} + K_{XX} \Box \pi) \nabla^\mu \pi \nabla^\nu \pi 
- 2(\nabla^\mu K_X) \nabla^\nu \pi - 2K_X \nabla^\mu \nabla^\nu \pi]
 \nabla_\mu \nabla_\nu \chi &
\nonumber\\
 + 2 K_X R_{\mu \nu}^{(1)} \nabla^\mu \pi \nabla^\nu \pi + \ldots &= 0 \; ,
\label{oct23-15-1} 
\end{align}
where $\pi$ is the background, $\chi$ is the
Galileon perturbation about this background,
and $R_{\mu \nu}^{(1)}$ is linear in metric perturbations.
We now make use of the Einstein equations $R_{\mu \nu} = 
\kappa \left( T_{\mu \nu} - \frac{1}{2} g_{\mu \nu} T^\lambda_\lambda \right) \;$, linearize the energy-momentum tensor \eqref{emt} and obtain for the last term 
in eq.~\eqref{oct23-15-1}
\[
 2 K_X R_{\mu \nu}^{(1)} \nabla^\mu \pi \nabla^\nu \pi
= - 2 \kappa K_X^2 \left[-X^2 \Box \chi + 4X
\nabla^\mu \pi \nabla^\nu \pi \nabla_\mu \nabla_\nu \chi \right] + \ldots \; .\]
The resulting linearized Galileon field equation is obtained from the following quadratic Lagrangian
\begin{equation} \label{L^2}
    \begin{aligned}
    L^{(2)} &= [F_X  + K_X \Box \pi - K_\pi +  \nabla_\nu (K_X \nabla^\nu \pi)]
     \nabla_\mu \chi \nabla^\mu \chi
    \\
    &+ [2 (F_{XX} + K_{XX} \Box \pi) \nabla^\mu \pi \nabla^\nu \pi 
    - 2(\nabla^\mu K_X) \nabla^\nu \pi - 2K_X \nabla^\mu \nabla^\nu \pi]
     \nabla_\mu \chi \nabla_\nu \chi
    \\
    & - k K_X^2 X^2 \nabla_\mu \chi
    \nabla^\mu \chi + 4k K_X^2 X
    \nabla^\mu \pi \nabla^\nu \pi \nabla_\mu \chi \nabla_\nu \chi \;,
    \end{aligned}
\end{equation}
where $K_{X} = \partial{K}/\partial{X}$, etc. 

Now we specify a background to be dynamical and spherically symmetric. The background metric in $4$-dimensional space-time can be written in the following form
\begin{equation} \label{metric}
    ds^2 = a^2 (r,t) dt^2 -  b^2 (r,t) dr^2 - c^2(r,t) \gamma_{\alpha \beta} dx^\alpha dx^\beta
    \; ,
\end{equation}
where $x^\alpha$ and $\gamma_{\alpha \beta}$ are coordinates and
metric on the unit $2$-dimensional sphere. The background scalar field $\pi(r,t)$ is also dynamical and spherically symmetric. In terms of the background metric \eqref{metric}, the quadratic action \eqref{L^2} takes the form
\begin{equation}
    S^{(2)} =\int \text{d}^4 x\,\sqrt{-g}\, \Bigg[ a^{-2} {\mathcal{K}}^{00} \dot{\chi}^2 - (ab)^{-1} {\mathcal{K}}^{t r} \dot{\chi} \chi' - b^{-2} {\mathcal{K}}^{rr} (\chi')^2 - c^{-2} {\mathcal{K}}^\Omega \gamma^{\alpha \beta} \partial_\alpha \chi \partial_\beta \chi + \dots \Bigg],
\end{equation}
where the omitted terms contain fewer derivatives of $\chi$, dot and prime denote derivatives w.r.t. time and radial coordinate. The coefficients $\mathcal{K}$ are given in the Appendix A. Then upon switching to the Fourier space we derive the dispersion relation which reads
\begin{equation}
    {\mathcal{K}}^{00} {\omega}^{2} = {\mathcal{K}}^{rr} {k_r}^2 + {\mathcal{K}}^{\Omega} {k_{\phi}}^2 + {\mathcal{K}}^{tr} \omega {k_r}.
\end{equation}
The necessary conditions for the absence of gradient instabilities, for high momenta are as follows:
\begin{equation}
    {\mathcal{K}}^{tr} {k_r}^2 + 4 {\mathcal{K}}^{00} ({\mathcal{K}}^{rr} {k_r}^2 + {\mathcal{K}}^{\Omega} {k_{\phi}}^2) \geq 0.
\end{equation}
Since $k_r$ and $k_\phi$ are independent, the necessary set of conditions for the absence of gradient instabilities has the form
\begin{equation}
\begin{cases}
    {\mathcal{K}}^{\Omega} {\mathcal{K}}^{00} \geq 0 \\ {\mathcal{K}}^{rr} \geq -\frac{ ({\mathcal{K}}^{tr})^2}{4 {\mathcal{K}}^{00}}
\end{cases}.
\end{equation}

Additionally we require $\mathcal{K}^{00}>0$ to ensure that the scalar perturbation is not a ghost. The speeds of propagation of perturbations in radial and angular directions are
\begin{equation}\label{snd_spd1}
    c_{r} = \frac{\partial \omega}{\partial k_{r}} = \frac{a}{b} \left( \frac{{\mathcal{K}} ^{tr}}{ {\mathcal{K}}^{00}}
    \pm \frac{\sqrt{({\mathcal{K}}^{tr})^2 + 4 {\mathcal{K}}^{00}{\mathcal{K}}^{rr}}}{2 {\mathcal{K}}^{00}}\right),
\end{equation}
\begin{equation}\label{snd_spd2}
    c_{\phi} = \frac{\partial \omega}{\partial k_{\phi}} = \frac{a}{c}\left( \sqrt{\frac{ {\mathcal{K}} ^ {\Omega}}{ {\mathcal{K}}^{00}}}\right).
\end{equation}

The set of necessary stability conditions for the high momentum regime, specifically the absence of ghosts and gradient instabilities is as follows
\begin{subequations} \label{stb_cnd}
    \begin{align}
        \label{G00>0} &{\mathcal{K}}^{00} > 0,\\
        \label{Gomega>0} &{\mathcal{K}}^{\Omega} \geq 0,\\
        \label{Grr>} &{\mathcal{K}}^{rr} \geq -\frac{ ({\mathcal{K}}^{tr})^2}{4 {\mathcal{K}}^{00}}.
    \end{align}
\end{subequations}
Here we do not study stability for low-momenta perturbations (tachyonic instabilities).

\section{No-go theorem.}
Our purpose is to extend the stability analysis to the dynamical case and check whether there could be a stable solution with metric \eqref{metric} and scalar field $\pi(r,t)$ within the cubic subclass of Horndeski theory \eqref{Lagrangians}. The argument of the no-go theorem in the static case is quite technical, and we show that it does not cover the dynamical case. For more clarity, we present here the proof for the static and cosmological no-go theorems.

\subsection{Static background.} \label{sec:static no-go}
\hspace{\parindent}We follow the method shown in \cite{1601R} to construct an argument for the no-go theorem in the static case. The main idea is to show that the manually introduced variable $\mathcal{Q}$ is always singular when the stability condition \eqref{G00>0} is satisfied.

The background metric was introduced above in \eqref{metric} but here we consider its static limit. The Galileon field is also dynamical and spherically symmetric $\pi = \pi (r)$. In the context of a Lorentzian wormhole which we consider in this section the coordinate $r$ runs from $-\infty$ to $+\infty$ and the metric coefficients are strictly positive and bounded from below:
\begin{equation} \label{restrictions}
    a(r) \geq a_{min} > 0 \; , \;\;\;\; b(r) \geq b_{min} > 0\;, \;\;\;\; c(r) \geq R_{min} > 0\; ,
\end{equation}
the $R_{min}$ here is the effective radius of the throat. The metric coefficients imply the following asymptotic behavior for the spatial asymptotically flat wormhole
\begin{equation} \label{assymptotics}
     a(r) \to a_{\pm} \; , \;\;\;\; b(r) \to b_{\pm} \;,  \;\;\;\;
    c(r) \to \pm r \; , \;\;\;\; \mbox{as} \;\; r\to \pm \infty \; ,
\end{equation}
where $a_{\pm}$ are positive constants. 

We use the static limit of the expressions for $\mathcal{K}^{0 0}$ \eqref{Eff_metric_G00} and the Einstein equations to obtain the following relation

\begin{equation}\label{no-go r static}
\frac{2}{a b} \pi^{\prime 2} \mathcal{K}^{00}(r)=-\mathcal{Q}^{\prime}-\frac{a b}{2}k \mathcal{Q}^2.
\end{equation}

The expression for $\mathcal{Q}$ reads

\begin{equation}\label{Q_r}
    \mathcal{Q}=\frac{1}{c}\left(2 \frac{c}{a b^3} K_X \pi^{\prime 3}+\frac{2}{k} \frac{c^{\prime}}{a b}\right).
\end{equation}

In \eqref{no-go r static} we used the static limit of the combination of Einstein equations that generally has the form  
\begin{equation} \label{Ei_eq_comb}
    T^0_0 - T^r_r = -\frac{2}{k c}\left(\frac{a}{b} \left(\frac{c^{\prime}}{{ab}}\right)^{\prime}+\frac{b}{a} \frac{\partial}{\partial{t}}\left(\frac{\dot{c}}{{ab}}\right)\right).
\end{equation}

Now we return to \eqref{no-go r static} and use the static limit of the stability condition \eqref{G00>0} to get the following inequality

\begin{equation} \label{before_int static}
    \frac{d}{dr}\frac{1}{\mathcal{Q}} > \frac{a b}{2} k > C,
\end{equation}

where C is a positive constant and here we have implied the conditions \eqref{restrictions}. Let us consider the whole interval where $\mathcal{Q} \neq 0$, which is limited by points $r_{\pm}$ where $\mathcal{Q} = 0$ or $r_{\pm} = \infty$. Therefore, $\mathcal{Q}^{-1} \rightarrow -\infty$ on the left boundary ($r\rightarrow r_{-}$) and $\mathcal{Q}^{-1} \rightarrow \infty$ on the right boundary ($r \rightarrow r_{+}$). Eventually, the variable $\mathcal{Q}^{-1}$ has to cross zero at some point, therefore $\mathcal{Q}$ must become infinitely large. We do not consider here the case of $\mathcal{Q}=0$ everywhere (Cuscuton), in the cubic subclass of Horndeski theory it leads to the relation $K^{00}=0$, hence the sound speeds \eqref{snd_spd1},\eqref{snd_spd2} become infinitely large. Cuscuton, on the other hand, is a subject that has been extensively researched \cite{Cuscuton}.

We also present here an alternative method for deriving the no-go theorem, which will be beneficial in further analysis. After integration from $r$ to $r^{\prime}>r$ the \eqref{before_int static} reads as
\begin{equation} \label{after_int static}
    \mathcal{Q}^{-1}(r) - \mathcal{Q}^{-1}(r^\prime) < (r^\prime - r) C.
\end{equation}
Suppose now that $\mathcal{Q}(r^\prime) > 0$ at some value of $r^\prime$. As $r$ decreases from $r^\prime$ to $-\infty$, $\mathcal{Q}^{-1}(r)$ starts positive, stays bounded from above and finally becomes bounded by a negative number. Hence, $\mathcal{Q}^{-1}(r)$ crosses zero, implying that
that $\mathcal{Q}$ becomes infinite and the configuration is singular.

\subsection{Dependence on time coordinate.}\label{sec:Dependence on t.}
\hspace{\parindent}Now we briefly discuss another particular case - cosmological limit of the background \eqref{metric}. The case of only time-dependent metric functions $a(t)$, $b(t)$ and the scalar field $\pi(t)$ also has a no-go theorem. Note that the metric function $c(r,t)$ can depend on both variables.

\begin{equation}\label{no-go t}
    2\frac{c}{a b}{\left(\dot{\pi}\right)}^{2} \mathcal{K}^{rr} = -\frac{\partial}{\partial{t}}\left(\frac{2c}{a^3b}K_X{\left(\dot{\pi}\right)}^{3} - \frac{2}{k}\frac{\dot{c}}{{ab}} \right) + \: \frac{k}{a^2} Kx {\left(\dot{\pi}\right)}^{3} \left(\frac{2c}{a^3b} Kx {\left(\dot{\pi}\right)}^{3} - \frac{2}{k}\frac{\dot{c}}{ab}\right),
\end{equation}

The new variable analogous to \eqref{Q_r} is introduced as follows

\begin{equation}\label{Q_t}
    \mathcal{O}=\frac{1}{c}\left(\frac{2c}{a^3b}K_X{\left(\dot{\pi}\right)}^{3} - \frac{2}{k}\frac{\dot{c}}{{ab}} \right).
\end{equation}

It casts the relation \eqref{no-go t} into

\begin{equation}\label{no-go t2}
\frac{2}{a b} {\dot{\pi}}^{2} \mathcal{K}^{rr}(r,t)=-\dot{\mathcal{O}}-\frac{a b}{2} k \mathcal{O}^2.
\end{equation}

Applying the stability condition \eqref{Grr>} we obtain
\begin{equation} \label{before_int_t}
    \frac{d}{dt}\frac{1}{\mathcal{O}} > \frac{a b}{2} k > C.
\end{equation}

The proof of this no-go theorem \cite{Cosmologies:2016} is similar to the static case and we conclude that $\mathcal{O}$ is singular at some point.

\subsection{Dependence on both variables.} \label{sec:Dependence on both variables}
Now we return to the general spherically symmetric, dynamical case, where all metric functions and the scalar field depend on both t and r. Then \eqref{no-go r static} modifies as follows
 
\begin{equation}\label{no-go r}
\frac{2}{a b} \pi^{\prime 2} \mathcal{K}^{00}(r,t)=-\mathcal{Q}^{\prime}-\frac{a b}{2}k \mathcal{Q}^2 + \mathcal{Y}(r,t).
\end{equation}

 Here we introduced the additional function $\mathcal{Y}$ \textbf{which is zero in the case of static scalar field}, its explicit form can be found in the Appendix B. We write the equation \eqref{no-go r} in this particular way because of its similarity to the static case. We will use it to check if the no-go theorem holds.

After use of the stability condition \eqref{G00>0}, from \eqref{no-go r} we obtain

\begin{equation} \label{before_int}
    \frac{d}{dr}\frac{1}{\mathcal{Q}} > \frac{a b}{2} k - \frac{\mathcal{Y}(r,t)}{\mathcal{Q}^2} > C - \frac{\mathcal{Y}(r,t)}{\mathcal{Q}^2},
\end{equation}

Naively, the problem which leads to the no-go theorem can be solved in the \textbf{dynamical} case by choosing the function $\mathcal{Y}$ in such a way as to make the right-hand side of \eqref{before_int} strictly negative for any moment of time, which breaks the argument of the static no-go theorem.

It can be shown in another manner. After integration from $r$ to $r^{\prime}>r$ the \eqref{before_int} reads as

\begin{equation} \label{after_int_dynamic}
    \mathcal{Q}^{-1}(r, t) - \mathcal{Q}^{-1}(r^\prime, t) < \int^{r^\prime}_{r}{\frac{\mathcal{Y}(r,t) - C \mathcal{Q}^2}{\mathcal{Q}^2}}{\mathrm{d}r}.
\end{equation}

Suppose that for some $r^{\prime}$, $\mathcal{Q}(r^{\prime},t)>0$ then for each fixed value of $t$ the integral on the right-hand side of the inequality \eqref{after_int_dynamic} is positive for $r \rightarrow -\infty$ and $\mathcal{Q}^{-1}(r,t)$ is bounded from above by a positive number, opposite to the static limit where it is bounded by a negative number. If $\mathcal{Q}$ is negative for some value of $r$, according to the constraint \eqref{after_int_dynamic}, it could remain negative for $r \rightarrow +\infty$.

On the other hand, in the preceding analysis we have considered only the stability condition \eqref{G00>0}, so, in principle, a different no-go theorem may reappear if more necessary restrictions \eqref{stb_cnd} are imposed on the system. We study the mentioned possibility in the Appendix B.

Extending the cosmological case to the time and radial dependent background, the \eqref{no-go t2} 
modifies as follows

\begin{equation}\label{no-go rt2}
\frac{2}{a b} {\dot{\pi}}^{2} \mathcal{K}^{rr}(r,t)=-\dot{\mathcal{O}}-\frac{a b}{2} k \mathcal{O}^2 + \mathcal{J}(r,t).
\end{equation}
the explicit expression for $\mathcal{J}(r,t)$ can be found in the Appendix B. Here we introduced the additional function $\mathcal{J}$ which is zero in the case of only time-dependent scalar field. The proof of the no-go theorem breaks for the same reason as in the static case extension.

\section{No-go theorem for a weak dependence on time}\label{sec:no-go weak on t}
\label{sec:no-go_ghosts}

\hspace{\parindent} Let us consider a scenario with a weak time dependence for $t \in G$ (where G is any set of time values) and for any r. We are going to use the fact that the additional function \eqref{Y_full} $\mathcal{Y}(r,t) = 0$ in the case of the static Galileon field $\pi(r,t) = \pi(r)$. We assume that in the region $t \in G,\; \forall r$ it is possible to split $\pi(r,t)$ as

\begin{equation} \label{pi_split}
\pi(r,t) = \tilde{\pi}(r) + \phi(r,t), \quad \frac{\dot{\phi}(r,t)}{\pi^\prime(r,t)}\ll 1.
\end{equation}

Then we apply the decomposition \eqref{pi_split} of the scalar field only for the additional function $\mathcal{Y}(r,t)$

\begin{equation}\label{Y_split} 
\mathcal{Y}(r,t) = \varepsilon(r,t), \quad t \in G, \quad \forall r.
\end{equation}

The notation $\varepsilon(r,t)$ corresponds to the "small" function, in \eqref{Y_split} we have used the fact that all terms of $\mathcal{Y}$ contain time derivatives of background scalar field. 

The absence of ghost instabilities in the dynamical case is provided by \eqref{G00>0}. Now we can use \eqref{pi_split} and \eqref{Y_split} to turn the right hand side of \eqref{no-go r} into:
\begin{equation}\label{small_Y}
    \frac{2}{a b} \pi^{\prime 2} \mathcal{K}^{00}(r,t)=-\mathcal{Q}^{\prime}-\frac{a b}{2} k \mathcal{Q}^2 + \varepsilon(r,t).
\end{equation}

The left hand side of \eqref{small_Y} is strictly positive, there we do not separate weak time dependence. In general $\mathcal{\tilde{Q}}\neq0$ and due to \eqref{G00>0}

\begin{equation} \label{weak r ineq1}
    \frac{d}{dr}\frac{1}{\mathcal{Q}} > \frac{a b}{2} k - \frac{\varepsilon(r,t)}{\mathcal{Q}^2} > C,
\end{equation}

We assume that the metric function $a(r,t)$ and $b(r,t)$ are bounded from below by positive numbers. The last term of \eqref{weak r ineq1} is of the next order in comparison to the first one and then the right-hand side of \eqref{weak r ineq1} is strictly positive and bounded from below by some positive constant $C$. As a result, the relation \eqref{weak r ineq1} implies that
\begin{equation}\label{weak r ineq2}
    \frac{d}{dr}\frac{1}{\mathcal{Q}} > C.
\end{equation}

We obtain an inequality similar to the \eqref{before_int static}, hence $\mathcal{Q}$ should become singular at some $r$. Note that the argument works only if time dependence is weak in the considered time region for {\it any} radial coordinate. Otherwise the argument breaks -- the reason for the statement was discussed in Sec.\ref{sec:Dependence on both variables}.


Above, we assumed $\mathcal{Q}\neq 0$. Now, if $\mathcal{Q}=0$, the argument is not applicable and the equality \eqref{small_Y} takes the following form
    \begin{equation}\label{lin_stb_cnd}
        \frac{2}{a b} \pi^{\prime 2} \mathcal{K}^{00}(r,t)=\varepsilon(r,t) > 0.
    \end{equation}

    The main difference from the static background case is in the presence of dynamical perturbations. Due to the weak dependence on the time coordinate, we can omit all terms that contain higher orders of small functions, and \eqref{lin_stb_cnd} can be satisfied by choosing $\varepsilon$. However, in the cases of small ${\mathcal{K}}^{00}$, ${\mathcal{K}}^{rr}$ is of zero order and generally remains finite, hence the sound speeds \eqref{snd_spd1} and \eqref{snd_spd2} become infinitely large.
    
\section{No-go theorem for weak dependence on radial coordinate}\label{sec:no-go weak on r}
    In analogy to Sec.\ref{sec:no-go_ghosts} one can extend the cosmological no-go theorem to the case of weak scalar field dependence on the radial coordinate in the region $r \in G,\; \forall t$. However, the difference lies in the condition \eqref{Grr>} which has the modified right-hand side, so we additionally assume that the coefficient $a(r,t)$ also has weak radial dependence

    \begin{subequations}
    \begin{align}\label{decomp_f_t}
    \pi(r,t) &= \tilde{\pi}(t) + \phi(r,t), \quad \frac{\phi^\prime}{\dot{\pi}}\ll1,\\
    a(r,t) &= \tilde a(t)+\alpha(r,t), \quad \frac{\alpha^\prime}{\dot{a}}\ll1.
    \end{align} 
    \end{subequations}

Then we apply the decomposition \eqref{decomp_f_t} to the functions $\mathcal{J}(r,t)$ and $\mathcal{K}^{tr}(r,t)$
    
    \begin{subequations}
    \begin{align}
    \mathcal{J}(r,t) &= \varepsilon_{1}(r,t),\label{J_split} \\
    \mathcal{K}^{tr}(r,t) & = \varepsilon_2(r,t). \label{K_split}
    \end{align} 
    \end{subequations}

The notations $\varepsilon_i(r,t)$ are for the "small" functions, which are the functions of $\phi(r,t), \; \alpha(r,t)$ and their derivatives. In \eqref{J_split} and \eqref{K_split} we have used the fact that all terms of $\mathcal{J}$ and $\mathcal{K}^{tr}$ contain time derivatives of background functions $\pi(r,t)$ or $a(r,t)$ which have a weak dependence on time.

The absence of gradient instabilities in the general case is provided by \eqref{Grr>}
    
\begin{equation}
    {\mathcal{K}}^{rr} \geq -\frac{ ({\mathcal{K}}^{tr})^2}{4 {\mathcal{K}}^{00}} = -\frac{\varepsilon_2(r,t)^2}{4 {\mathcal{K}}^{00}}.
\end{equation} 

Whereupon the relation \eqref{no-go rt2} turns into

\begin{equation}\label{no-go rt2 weak}
\frac{2 }{a b} {\dot{\pi}}^{2} \mathcal{K}^{rr}(r,t)=-\dot{\mathcal{O}}-\frac{a b}{2} k \mathcal{O}^2 + \mathcal{\varepsilon}_1(r,t) \geq - \frac{2}{a b} {\dot{\pi}}^{2} \cdot \frac{\varepsilon_2(r,t)^2}{2 {\mathcal{K}}^{00}}.
\end{equation}

Then we cast the \eqref{no-go rt2 weak} into

\begin{equation}
    \frac{d}{dt}\frac{1}{\mathcal{O}} \geq \frac{a b}{2} k - \frac{\varepsilon_1(r,t)}{\mathcal{O}^2} - \frac{2}{a b} {\dot{\pi}}^{2} \cdot \frac{\varepsilon_2(r,t)^2}{2 {\mathcal{K}}^{00}} > C,
\end{equation}

where $C$ is a positive constant. We use the assumption \eqref{decomp_f_t} to neglect the $\varepsilon_i$-terms, so the theorem returns to the cosmological no-go theorem \ref{sec:Dependence on t.}.

Sections \ref{sec:no-go weak on t} and \ref{sec:no-go weak on r} show that any stable non-singular solution should not have regions with a weak scalar field dependence on any variable, time or radial. Common compact objects in an expanding universe have asymptotic regions with weak radial coordinate dependence, which is commonly used as a spatial asymptotic of function behavior. One of the ways to avoid the no-go for compact objects is to make partial derivatives of the scalar field oscillate similarly in space and time directions. Below we refer to both theorems related to the weak dependence as weak no-go theorems.

\section{Generalized no-go theorem}

We note that the additional functions $\mathcal{Y}(r,t)$ and $\mathcal{J}(r,t)$, as was mentioned before, vanish in the cases of only time- and radial-dependent scalar field respectively. In the present section, we study the possibility of setting these functions to zero through a redefinition of coordinates. 
After making a coordinate redefinition $(r,t) \rightarrow (\tilde{r}, \tilde{t})$, let us start with the radial no-go theorem. We can one more time derive the set of stability conditions, and the expression \eqref{after_int_dynamic} from the radial no-go theorem modifies in the new coordinate patch as follows:

\begin{equation}
    \mathcal{Q}^{-1}(\tilde{r}, \tilde{t}) - \mathcal{Q}^{-1}(\tilde{r}^\prime, \tilde{t}) < -C(\tilde{r}^\prime - \tilde{r}) + \int^{\tilde{r}^\prime}_{\tilde{r}}{\frac{\mathcal{\tilde{Y}}(\tilde{r},\tilde{t})}{\mathcal{Q}^2}}{\mathrm{d}\tilde{r}}.
\end{equation}

The main idea is to choose new coordinates in such a way that the scalar field $\pi(r,t)$ becomes a function with a weak dependence on the time or radial coordinate to reduce the case to the previous no-go theorems. To accomplish this, there must be a curve $\gamma(\lambda)$ a surface of constant angles in (r,t) surface (submanifold) whose tangent vector $\xi^{\mu}(\lambda) = \frac{\partial x^\mu}{\partial\lambda}$ is always close to the gradient of the scalar field $\partial_{\mu} \pi(r,t)$, while the normal vector $n^{\mu}(\lambda)=\frac{\partial^2 x^\mu}{\partial\lambda^2}/\left|\frac{\partial^2 x^\mu}{\partial\lambda^2}\right|$ should locally match the contour lines of the scalar field in the region around the curve $\gamma$. In other words, we require $g^{\mu \nu} n_{\mu}\partial_{\nu}\pi < \varepsilon$. Then, in the local region around the curve, we perform a coordinate transformation to the Frenet basis. The coordinate transformation must be diffeomorphic and result in a background metric of the form \eqref{metric}. We use the notation $\varepsilon$ as a parameter that designs a weakness of the dependence of the scalar field on the second variable ($\tilde{r}$ or $\tilde{t}$) in the new basis.

The chosen curve $\gamma$ should be either timelike or spacelike to avoid changes in the metric signature. Due to the existence of both radial and time no-go theorems, both types of curves are allowed. To reduce a case to the no-go theorem with weak time dependence, the curve $\gamma$ should be spacelike ($\tilde{r}=\lambda$), and for the weak radial dependence, it should be timelike ($\tilde{t}=\lambda$). As the tangent vector of the curve $\gamma$ is defined by the gradient of the scalar field $\pi(r,t)$, the existence of this curve within a specified background depends only on the particular form of the scalar field and does not depend on the form of the metric functions. Finding an ansatz for the scalar field that violates these conditions is challenging, and it remains uncertain whether it can be achieved.

The important remark needs to be made for the generalized theorem in the case of reducing to the weak radial dependent scalar field (generalized cosmological no-go theorem). For general background, the stability condition \eqref{Grr>} has a modified r.h.s. which couldn't be set to zero by coordinate transformation at the same time as $\mathcal{J}$, due to its explicit form App. A \eqref{Ktr}. The possible way to obtain the no-go theorem is to restrict the background to be spatially asymptotically flat. To be more precise, it suffices to demand $\mathcal{K}^{tr} \rightarrow 0$, and by applying the explicit expression \eqref{Ktr} it results in a sufficient condition $\partial_{r}{a(r,t)} \rightarrow 0, \, r \rightarrow \pm \infty$. Here we always keep the perturbations dynamic, $\mathcal{K}^{00}>0$, to prevent singularities in the speeds \eqref{snd_spd1} and \eqref{snd_spd2}. Thus, as the radial coordinate takes high values $r\rightarrow \pm \infty$, the right-hand side of \eqref{Grr>} will tend to zero, recovering the validity of the generalized no-go theorem. The no-go theorem is now applicable to scenarios involving compact objects (possibly singular) within an expanding universe.

We summarize the sufficient conditions for the generalized no-go theorems once again below.
\hfill \break

$\exists \gamma(\lambda) \in (r,t)$: at any point of $\gamma$, its normal vector $n^{\mu}(\lambda)=\frac{\partial^2 x^\mu}{\partial\lambda^2}/\left|\frac{\partial^2 x^\mu}{\partial\lambda^2}\right|$  satisfies $g^{\mu \nu} n_{\mu}\partial_{\nu}\pi < \varepsilon$ and the following conditions are met in the region around the curve:
    \begin{enumerate}
        \item The stability conditions are satisfied.
        \item The field equations are satisfied.
        \item The curve $\gamma$ is either timelike or spacelike.
        \item $\gamma$ does not contain singularities of the background functions and the restrictions \eqref{restrictions} are met.
    \end{enumerate}
        \hfill \break
    As a result:
         \begin{itemize}
    \item if  $\gamma$ is spacelike then $\pi(\tilde{r},\tilde{t})\mid_{\gamma} = \pi(\tilde{r}) + \phi(\tilde{r},\tilde{t})$, $\frac{\dot{\phi}}{\pi^\prime}\ll1$, hence $\tilde{\mathcal{Y}}(\tilde{r},\tilde{t})\mid_{\gamma} = \varepsilon_1(\tilde{r},\tilde{t}) \Rightarrow$ case reduces to the weak radial no-go theorem.
    \item if $\gamma$ is timeelike then $\pi(\tilde{r},\tilde{t})\mid_{\gamma} = \pi(\tilde{t}) + \phi(\tilde{r},\tilde{t})$, $\frac{\phi^\prime}{\dot{\pi}}\ll1$, hence $\tilde{\mathcal{J}}(\tilde{r},\tilde{t})\mid_{\gamma} = \varepsilon_1(\tilde{r},\tilde{t}) \Rightarrow$ case reduces to the cosmological weak cosmological no-go theorem with a proviso that $a^{\prime}(r,t) \rightarrow 0, \; r\rightarrow \pm \infty$
    \end{itemize}

The proposed method shares certain similarities with the choosing of unitary gauge in the ADM formalism \cite{BH_Langlois2014}. Note, however, that in our case there is no global restriction on the scalar field, while the unitary gauge requires everywhere a time-like gradient of the scalar field. Our requirement, if any, for the gradient of the scalar field is to be time-like or space-like on a \textbf{particular curve $\gamma$} only, which is definitely an integration curve in the preceding theorem. Moreover, all sufficient conditions of the generalized no-go theorem are imposed on the background functions only in the region around the particular curve $\gamma$, so it is a distinctive feature of this theorem.

Below we consider an example of a scalar field that does not satisfy the conditions of the weak no-go theorem, but the generalized no-go theorem is applicable. The ansatz for the scalar field has the form
\begin{equation}\label{pi_cosh}
    \pi(r,t) = \cosh{(r t)}.
\end{equation}

\begin{figure} 
[H]\begin{center}\hspace{-1cm}
{\includegraphics[width=0.5\textwidth] {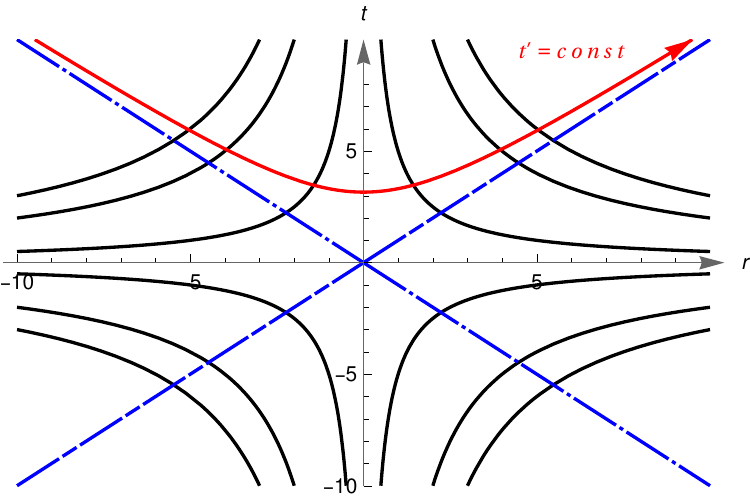}}

\caption{\footnotesize{The black lines are contour lines of the scalar field that were projected on the coordinate plane. The blue dashed lines corresponded to the light cone. The red line is the gradient line which is the new integration curve.}} \label{f-g}
\end{center}\end{figure}

The proof is shown in the Figure \ref{f-g}, where we find one spacelike curve\footnote{This is valid for arbitrary metric functions, at least for large coordinate values, as the metric should be spatially asymptotically flat.} which is the gradient curve for the scalar field, for this reason it is possible to make a coordinate transformation to replace the integration along the radial axis with the integration along a selected curve on which the conditions of the no-go theorem \ref{sec:no-go weak on t} are met.

\section{Conclusion}

\hspace{\parindent}In this work, we have analyzed the cubic subclass of Horndeski theory using the DPSV approach \cite{DPSV} and derived the stability conditions for high-momenta modes about dynamic, spherically symmetric background. This approach does not allow one to get a full set of stability conditions including tachyonic instabilities (low-momenta regime), but the conditions for high momentum and frequencies were derived, specifically the conditions for absence of ghost and gradient instabilities, and also the propagation speeds of perturbations were obtained. In the cubic subclass of Horndeski theory the no-go theorems exist for static spherically-symmetric backgrounds and cosmological scenarios \cite{1601R, Cosmologies:2016}, we have generalized it to backgrounds with weak dependence on the radial or time coordinate. It is clear that in many physically viable solutions there should be regions, especially asymptotic, that satisfy the conditions of the generalized no-go theorem. For example, a star in the expanding universe will have weak radial dependence in some region for any time, and it is one appropriate case for applying the no-go theorem for weak radial dependence. Even though, it was shown that the proof of the no-go theorem is not straightforwardly applicable to the general dynamical case, we have shown a possible way to bypass the mathematical argument of the no-go theorem in the dynamical case. However, constructing a stable solution encounters difficulties due to the lack of arbitrary functions to build a solution without solving the PDE system \eqref{CombNoKF}.

We have also derived the further step of generalizing the no-go theorems which allows us to reduce the dependence of the scalar field to one variable. The generalization of the spherically symmetric no-go theorem includes the wide class of backgrounds that have mild restrictions only on the scalar field (excluding only exotic examples) and lets all metric functions be fully arbitrary.

The extension of the cosmological no-go theorem arises under the condition that the background should be asymptotically flat. Now both theorems apply not only to non-singular solutions: cases such as bouncing universe with a black hole are also restricted. The sufficient conditions of the no-go theorem include the presence of either a timelike or a spacelike gradient curve which does not contain a scalar field singularity, so it does not contradict the fact of the existence of a singularity in the whole system. Our constraints on the scalar field are considerably less stringent compared to the global necessity for a timelike gradient of the scalar field, required to uphold the unitary gauge in the ADM formalism. These restrictions make almost any ansatz for the scalar field and asymptotically flat metric functions on spatial infinity corresponding to a compact object or a cosmological solution incompatible with the stability conditions in the cubic subclass of Horndeski theory, regardless of the presence of a singularity.

\section*{Acknowledgements}
The authors are grateful to A. Shtennikova, V. Volkova, Y. Ageeva for valuable discussions. The authors wish to thank Kasper
Peeters for developing and maintaining cadabra2 software \cite{cadabra}, with which most of the calculations
were performed. The work on this project has been supported by Russian Science Foundation grant № 24-72-10110,
\href{https://rscf.ru/project/24-72-10110/}{  https://rscf.ru/project/24-72-10110/}.

\newpage
\section{Appendix}

\subsection{Effective Metric}\label{Apndx_A}
\begin{subequations} \label{Eff_metric}
     \begin{align}
    \label{Eff_metric_G00}
    \mathcal{K}^{0 0} &= F_{X}+ 2F_{X X} {\left(\dot{\pi}\right)}^{2} {a}^{-2} - K_{\pi} -4K_{X} {c}^\prime {\pi}^\prime {b}^{-2} {c}^{-1}+2K_{X} {b}^\prime {\pi}^\prime {b}^{-3}+4K_{X} \dot{c} \dot{\pi} {a}^{-2} {c}^{-1}\nonumber\\
    &+2K_{X} \dot{b} \dot{\pi} {a}^{-2} {b}^{-1} -2K_{X} {\pi}^{\prime\prime} {b}^{-2} +2K_{X X} {b}^\prime {\pi}^\prime {\left(\dot{\pi}\right)}^{2} {a}^{-2} {b}^{-3}-2K_{X X} \dot{b} \dot{\pi} {\left({\pi}^\prime\right)}^{2} {a}^{-2} {b}^{-3}-2K_{X X} {b}^\prime {\left({\pi}^\prime\right)}^{3} {b}^{-5}\nonumber\\
    &-4K_{X X} {c}^\prime {\pi}^\prime {\left(\dot{\pi}\right)}^{2} {a}^{-2} {b}^{-2} {c}^{-1}+4K_{X X} \dot{c} {\left(\dot{\pi}\right)}^{3} {a}^{-4} {c}^{-1}+2K_{X X} \dot{b} {\left(\dot{\pi}\right)}^{3} {a}^{-4} {b}^{-1} -2K_{X X} {\pi}^{\prime\prime} {\left(\dot{\pi}\right)}^{2} {a}^{-2} {b}^{-2} \nonumber\\
    &+2K_{X X} {\pi}^{\prime\prime} {\left({\pi}^\prime\right)}^{2} {b}^{-4} -K_{X \pi} {\left(\dot{\pi}\right)}^{2} {a}^{-2}-K_{X \pi} {\left({\pi}^\prime\right)}^{2} {b}^{-2}-{K_{X}}^{2} {\left({\pi}^\prime\right)}^{4} {b}^{-4} k\nonumber\\
    &-2{K_{X}}^{2} {\left({\pi}^\prime\right)}^{2} {\left(\dot{\pi}\right)}^{2} {a}^{-2} {b}^{-2} k +3{K_{X}}^{2} {\left(\dot{\pi}\right)}^{4} {a}^{-4} k ,\\
    \label{Ktr}
    \mathcal{K}^{t r} &= -4F_{X X} {\pi}^\prime \dot{\pi} {a}^{-2} {b}^{-2} -4K_{X} {a}^\prime \dot{\pi} {a}^{-3} {b}^{-2}-4K_{X} {\pi}^\prime \dot{b} {a}^{-2} {b}^{-3} +4K_{X} {\dot{\pi}}^{\prime} {a}^{-2} {b}^{-2}\nonumber\\
    &-4K_{X X} {a}^\prime {\left(\dot{\pi}\right)}^{3} {a}^{-5} {b}^{-2}+4K_{X X} \dot{b} {\left({\pi}^\prime\right)}^{3} {a}^{-2} {b}^{-5}-8K_{X X} {\pi}^\prime \dot{c} {\left(\dot{\pi}\right)}^{2} {a}^{-4} {b}^{-2} {c}^{-1}\nonumber\\
    &+8K_{X X} {c}^\prime \dot{\pi} {\left({\pi}^\prime\right)}^{2} {a}^{-2} {b}^{-4} {c}^{-1}-4K_{X X} {\dot{\pi}}^{\prime} {\left({\pi}^\prime\right)}^{2} {a}^{-2} {b}^{-4}+4K_{X X} {\dot{\pi}}^{\prime} {\left(\dot{\pi}\right)}^{2} {a}^{-4} {b}^{-2}\nonumber\\
    &+4K_{X X} {a}^\prime \dot{\pi} {\left({\pi}^\prime\right)}^{2} {a}^{-3} {b}^{-4}-4K_{X X} {\pi}^\prime \dot{b} {\left(\dot{\pi}\right)}^{2} {a}^{-4} {b}^{-3}\nonumber\\
    &+4K_{X \pi} {\pi}^\prime \dot{\pi} {a}^{-2} {b}^{-2} + 8\dot{\pi} {K_{X}}^{2} {\left({\pi}^\prime\right)}^{3} {a}^{-2} {b}^{-4} k-8{\pi}^\prime {K_{X}}^{2} {\left(\dot{\pi}\right)}^{3} {a}^{-4} {b}^{-2} k,\\
    \mathcal{K}^{r r} &= F_{X}-2F_{X X} {\left({\pi}^\prime\right)}^{2} {b}^{-2} -K_{\pi} \nonumber\\
    &-4K_{X} {c}^\prime {\pi}^\prime {b}^{-2} {c}^{-1}-2K_{X} {a}^\prime {\pi}^\prime {a}^{-1} {b}^{-2}+2K_{X} \ddot{\pi} {a}^{-2}+4K_{X} \dot{c} \dot{\pi} {a}^{-2} {c}^{-1}-2K_{X} \dot{a} \dot{\pi} {a}^{-3}\nonumber\\
    &+4K_{X X} {c}^\prime {\left({\pi}^\prime\right)}^{3} {b}^{-4} {c}^{-1}+2K_{X X} {a}^\prime {\left({\pi}^\prime\right)}^{3} {a}^{-1} {b}^{-4}-2K_{X X} \ddot{\pi} {\left({\pi}^\prime\right)}^{2} {a}^{-2} {b}^{-2}\nonumber\\
    &-4K_{X X} \dot{c} \dot{\pi} {\left({\pi}^\prime\right)}^{2} {a}^{-2} {b}^{-2} {c}^{-1}-2K_{X X} {a}^\prime {\pi}^\prime {\left(\dot{\pi}\right)}^{2} {a}^{-3} {b}^{-2}+2K_{X X} \dot{a} \dot{\pi} {\left({\pi}^\prime\right)}^{2} {a}^{-3} {b}^{-2}\nonumber\\
    &+2K_{X X} \ddot{\pi} {\left(\dot{\pi}\right)}^{2} {a}^{-4}-2K_{X X} \dot{a} {\left(\dot{\pi}\right)}^{3} {a}^{-5}+K_{X \pi} {\left({\pi}^\prime\right)}^{2} {b}^{-2}%
    +K_{X \pi} {\left(\dot{\pi}\right)}^{2} {a}^{-2}\nonumber\\
    &+3{K_{X}}^{2} {\left({\pi}^\prime\right)}^{4} {b}^{-4} k-2{K_{X}}^{2} {\left({\pi}^\prime\right)}^{2} {\left(\dot{\pi}\right)}^{2} {a}^{-2} {b}^{-2} k-{K_{X}}^{2} {\left(\dot{\pi}\right)}^{4} {a}^{-4} k,\\
    \mathcal{K}^{\Omega} &= F_{X} - 2K_{X} {\pi}^{\prime\prime} {b}^{-2}-2K_{X} {c}^\prime {\pi}^\prime {b}^{-2} {c}^{-1}+2K_{X} {b}^\prime {\pi}^\prime {b}^{-3}\nonumber\\
    &-2K_{X} {a}^\prime {\pi}^\prime {a}^{-1} {b}^{-2}+2K_{X} \ddot{\pi} {a}^{-2}+2K_{X} \dot{c} \dot{\pi} {a}^{-2} {c}^{-1}+2K_{X} \dot{b} \dot{\pi} {a}^{-2} {b}^{-1}-2K_{X} \dot{a} \dot{\pi} {a}^{-3}\nonumber\\
    &+2K_{X X} {\pi}^{\prime\prime} {\left({\pi}^\prime\right)}^{2} {b}^{-4}-2K_{X X} {b}^\prime {\left({\pi}^\prime\right)}^{3} {b}^{-5}-4K_{X X} {\pi}^\prime \dot{\pi} {\dot{\pi}}^{\prime} {a}^{-2} {b}^{-2}+2K_{X X} \dot{b} \dot{\pi} {\left({\pi}^\prime\right)}^{2} {a}^{-2} {b}^{-3}\nonumber\\
    &+2K_{X X} {a}^\prime {\pi}^\prime {\left(\dot{\pi}\right)}^{2} {a}^{-3} {b}^{-2}+2K_{X X} \ddot{\pi} {\left(\dot{\pi}\right)}^{2} {a}^{-4}-2K_{X X} \dot{a} {\left(\dot{\pi}\right)}^{3} {a}^{-5}-K_{X \pi} {\left({\pi}^\prime\right)}^{2} {b}^{-2}\nonumber\\
    &+K_{X \pi} {\left(\dot{\pi}\right)}^{2} {a}^{-2} -{K_{X}}^{2} {\left({\left({\pi}^\prime\right)}^{2} {a}^{2} -K_{\pi} -{\left(\dot{\pi}\right)}^{2} {b}^{2}\right)}^{2} {a}^{-4} {b}^{-4} k .
    \end{align}
\end{subequations}

\subsection{The possibility of algebraic generalization of no-go theorem}

\hspace{\parindent} We have already discussed two limits of the no-go theorem. The way of generalization should include both of them. In Sec.\ref{sec:Dependence on both variables} we see the lack of possibility of building an argument for the no-go theorem using only one stability condition. The purpose of this section is to find a more general combination and study whether the no-go theorem holds or not in the dynamical case.

In full analogy with $\mathcal{Y}$ in extension of the static no-go theorem \eqref{no-go r}, for the cosmological no-go theorem we introduce an additional function $\mathcal{J}(r,t)$ which definition has the following form

\begin{equation}\label{J}
\begin{aligned}
    \mathcal{J}(r,t) = \frac{2}{a^2}{\left(\dot{\pi}\right)}^{2} \mathcal{K}^{rr}+2 \frac{b}{a c}\frac{\partial}{\partial{t}}\left(\frac{c}{a^3 b} {\left(\dot{\pi}\right)}^{3} K_{X}\right)-k K_{X} {\left(\dot{\pi}\right)}^{3} \frac{b}{a^3 c} \left(\frac{2 c}{a^3 b} K_{X} {\left(\dot{\pi}\right)}^{3}-\frac{2}{k} \frac{\dot{c}}{a b}\right) +T_{t}^{t}-T_{r}^{r}.
\end{aligned}
\end{equation}

The explicit expressions for the additional functions read as

\begin{equation} \label{Y_full}
    \begin{aligned}
    \frac{b}{a}\mathcal{Y}(r,t) &= \ddot{\pi} K_{X} \left(-2 {\left({\pi}^\prime\right)}^{2} {a}^{-2} {b}^{-2}+2{\left(\dot{\pi}\right)}^{2} {a}^{-4}\right) \\
    & +\dot{\pi} \left[ -2F_{X} \dot{\pi} {a}^{-2}+4F_{X X} \dot{\pi} {\left({\pi}^\prime\right)}^{2} {a}^{-2} {b}^{-2}+2K_{\pi} \dot{\pi} {a}^{-2}+ K_{X} \left(2\dot{\pi} {\pi}^{\prime\prime} {a}^{-2} {b}^{-2}+4{c}^\prime {\pi}^\prime \dot{\pi} {a}^{-2} {b}^{-2} {c}^{-1} \right. \right. \\
    &\left. \left.+4\dot{c} {\left({\pi}^\prime\right)}^{2} {a}^{-2} {b}^{-2} {c}^{-1}-2{b}^\prime {\pi}^\prime \dot{\pi} {a}^{-2} {b}^{-3}+6\dot{b} {\left({\pi}^\prime\right)}^{2} {a}^{-2} {b}^{-3}-2{a}^\prime {\pi}^\prime \dot{\pi} {a}^{-3} {b}^{-2}+2\dot{a} {\left({\pi}^\prime\right)}^{2} {a}^{-3} {b}^{-2} \right. \right. \\
    &\left. \left. -4\dot{c} {\left(\dot{\pi}\right)}^{2} {a}^{-4} {c}^{-1}-2\dot{b} {\left(\dot{\pi}\right)}^{2} {a}^{-4} {b}^{-1}-2\dot{a} {\left(\dot{\pi}\right)}^{2} {a}^{-5} \right) +K_{X X} \left(4{\dot{\pi}}^{\prime} {\left({\pi}^\prime\right)}^{3} {a}^{-2} {b}^{-4}-4\dot{\pi} {\pi}^{\prime\prime} {\left({\pi}^\prime\right)}^{2} {a}^{-2} {b}^{-4} \right. \right.\\
    & \left. \left. -8{c}^\prime \dot{\pi} {\left({\pi}^\prime\right)}^{3} {a}^{-2} {b}^{-4} {c}^{-1}+4{b}^\prime \dot{\pi} {\left({\pi}^\prime\right)}^{3} {a}^{-2} {b}^{-5}-4\dot{b} {\left({\pi}^\prime\right)}^{4} {a}^{-2} {b}^{-5}-4{a}^\prime \dot{\pi} {\left({\pi}^\prime\right)}^{3} {a}^{-3} {b}^{-4}+8\dot{c} {\left({\pi}^\prime\right)}^{2} {\left(\dot{\pi}\right)}^{2} {a}^{-4} {b}^{-2} {c}^{-1} \right. \right.\\
    &\left. \left.+4\dot{b} {\left({\pi}^\prime\right)}^{2} {\left(\dot{\pi}\right)}^{2} {a}^{-4} {b}^{-3}\right)-2K_{X \pi} \dot{\pi} {\left({\pi}^\prime\right)}^{2} {a}^{-2} {b}^{-2} +{K_{X}}^{2} \left(-4\dot{\pi} {\left({\pi}^\prime\right)}^{4} {a}^{-2} {b}^{-4} k+6{\left({\pi}^\prime\right)}^{2} {\left(\dot{\pi}\right)}^{3} {a}^{-4} {b}^{-2} k\right) \right],
    \end{aligned}
\end{equation}

\begin{equation} \label{J_full}
    \begin{aligned}
    \frac{a}{b}\mathcal{J}(r,t) &= {\pi}^{\prime\prime} K_{X} \left(-2 {\left(\dot{\pi}\right)}^{2} {a}^{-2} {b}^{-2}+2 {\left({\pi}^\prime\right)}^{2} {b}^{-4}\right)\\
    &+ \pi^{\prime} \left[2F_{X} {\pi}^\prime {b}^{-2}+4F_{X X} {\pi}^\prime {\left(\dot{\pi}\right)}^{2} {a}^{-2} {b}^{-2} -2K_{\pi} {\pi}^\prime {b}^{-2} +K_{X} \left(-4{c}^\prime {\left({\pi}^\prime\right)}^{2} {b}^{-4} {c}^{-1}-2{b}^\prime {\left({\pi}^\prime\right)}^{2} {b}^{-5}\right. \right.\\
    &\left. \left.-2{a}^\prime {\left({\pi}^\prime\right)}^{2} {a}^{-1} {b}^{-4}+2\ddot{\pi} {\pi}^\prime {a}^{-2} {b}^{-2}+4{c}^\prime {\left(\dot{\pi}\right)}^{2} {a}^{-2} {b}^{-2} {c}^{-1}+4\dot{c} \dot{\pi} {\pi}^\prime {a}^{-2} {b}^{-2} {c}^{-1}+2{b}^\prime {\left(\dot{\pi}\right)}^{2} {a}^{-2} {b}^{-3}\right. \right.\\
    &\left. \left.-2\dot{b} \dot{\pi} {\pi}^\prime {a}^{-2} {b}^{-3}+6{a}^\prime {\left(\dot{\pi}\right)}^{2} {a}^{-3} {b}^{-2}-2\dot{a} \dot{\pi}{\pi}^\prime {a}^{-3} {b}^{-2}\right) - 2K_{X \pi} {\pi}^\prime {\left(\dot{\pi}\right)}^{2} {a}^{-2} {b}^{-2}\right.\\
    &\left.+K_{X X} \left(-8{c}^\prime {\left({\pi}^\prime\right)}^{2} {\left(\dot{\pi}\right)}^{2} {a}^{-2} {b}^{-4} {c}^{-1}-4{a}^\prime {\left({\pi}^\prime\right)}^{2} {\left(\dot{\pi}\right)}^{2} {a}^{-3} {b}^{-4}+4\ddot{\pi} {\pi}^\prime {\left(\dot{\pi}\right)}^{2} {a}^{-4} {b}^{-2}-4{\dot{\pi}}^{\prime} {\left(\dot{\pi}\right)}^{3} {a}^{-4} {b}^{-2}\right. \right.\\
    &\left. \left.+8\dot{c} {\pi}^\prime {\left(\dot{\pi}\right)}^{3} {a}^{-4} {b}^{-2} {c}^{-1}+4\dot{b} {\pi}^\prime {\left(\dot{\pi}\right)}^{3} {a}^{-4} {b}^{-3}+4{a}^\prime {\left(\dot{\pi}\right)}^{4} {a}^{-5} {b}^{-2}-4\dot{a} {\pi}^\prime {\left(\dot{\pi}\right)}^{3} {a}^{-5} {b}^{-2}\right) \right.\\
    &\left. +{K_{X}}^{2} \left(-6{\left({\pi}^\prime\right)}^{3} {\left(\dot{\pi}\right)}^{2} {a}^{-2} {b}^{-4} k+4{\pi}^\prime {\left(\dot{\pi}\right)}^{4} {a}^{-4} {b}^{-2} k\right) \right].
    \end{aligned}
\end{equation}

The following particular equality has a relatively compact form

\begin{equation}
\begin{aligned} \label{combYJ}
    2\dot{\pi} {\pi}^{\prime} \left( \frac{a}{b} \mathcal{Y} + \frac{b}{a} \mathcal{J} \right) =& -2 a^2 \dot{\pi} \frac{\partial}{\partial{r}}\left(K_{X}\dot{\pi}^2 (\pi^\prime)^2 \frac{1}{b^2 a^4}\right) + 2 b^2 \pi^\prime\frac{\partial}{\partial{t}}\left( K_{X}\dot{\pi}^2 (\pi^\prime)^2 \frac{1}{a^2 b
    ^4}\right)\\
    & +\frac{2}{3} \dot{\pi} {\pi}^{\prime} K_{X} c^6 \left[ \frac{a^3}{b} \frac{\partial}{\partial{r}}\left((\pi^\prime)^3 \frac{1}{a^3 b^3 c^6}\right) - \frac{b^3}{a} \frac{\partial}{\partial{t}}\left((\dot{\pi})^3 \frac{1}{a^3 b^3 c^6}\right) \right) \\
    & +2\left(\frac{(\pi^\prime)^2}{b^2} - \frac{\dot{\pi}^2}{a^2}\right) \left( -K_X^2 \frac{(\pi^\prime)^2 \dot{\pi}^2}{a^2 b^2}k - K_{\pi} + F_{X} \right).
\end{aligned}
\end{equation}

Generalization of the no-go theorem argument in the dynamical case with radial dependence is not straightforward due to non-linearity in the stability conditions \eqref{Grr>}.

\subsection{Solving the system of motion equations for background field}
\hspace{\parindent}It is possible to get a combination of Einstein equations that does not contain Lagrangian functions K, F and their derivatives

\begin{equation} \label{CombNoKF}
    \begin{aligned}
        S &= \frac{(\pi^\prime)^2}{a b} \left[ \left( \frac{a^2 c}{b} \left( \frac{a}{c}\right)^\prime \right)^\prime \frac{1}{c^2} + \frac{\partial}{\partial{t}}{\left( \frac{1}{b c} \frac{\partial}{\partial{t}}{\left(a c\right)} \right) } - \frac{2 \dot{a} \dot{c}}{b c} \right) \\
        &+\frac{(\dot{\pi})^2}{a b} \left[ \frac{\partial}{\partial{t}}{\left( \frac{a^2 c}{b} \frac{\partial}{\partial{t}}{\left( \frac{a}{c}\right)} \right)} \frac{1}{c^2} + \left( \frac{1}{b c} \left(a c\right)^\prime \right)^\prime - \frac{2 a^\prime c^\prime}{b c} \right)\\
        &+\frac{4 \dot{\pi} \pi^\prime}{b^2 c}\left(\dot{c^\prime} - \frac{\dot{b}}{b} c^\prime - \frac{a^\prime}{a} \dot{c} \right) 
    \end{aligned}
\end{equation}

This combination is similar to the one shown in \cite{2212wormhole} for $G_4 = 1$ and $F_4 = 0$, the absence of Lagrangian functions there is a feature of the cubic subclass of Horndeski theory.

The relation \eqref{CombNoKF} is the second-order nonlinear PDE for $a$, $b$, $c$, $\pi$ with their special asymptotics. The method of Lagrangian reconstruction that aims to evade PDEs is frequently used to obtain stable solutions in Horndeski theory \cite{1812wormhole, 2212wormhole}. However, in our case \eqref{CombNoKF} does not contain the Lagrangian functions, so it is impossible to avoid solving PDEs while constructing a solution in the cubic subclass of Horndeski theory opposite to the higher subclasses of it. The proof of absence of the no-go theorem should include the way to construct stable non-singular solution, and we do not provide it.


\begin{thebibliography}{}

\bibitem{Horndeski}
  G.~W.~Horndeski,
  ``Second-order scalar-tensor field equations in a four-dimensional space,''
  Int.\ J.\ Theor.\ Phys.\  {\bf 10} (1974) 363.

\bibitem{Trincherini:2008}
A.~Nicolis, R.~Rattazzi and E.~Trincherini,
``The Galileon as a local modification of gravity,''
Phys. Rev. D \textbf{79} (2009), 064036
\href{https://arxiv.org/pdf/0811.2197.pdf}{[arXiv:0811.2197 [hep-th]]}.

\bibitem{Deser:2009}
C.~Deffayet, S.~Deser and G.~Esposito-Farese,
``Generalized Galileons: All scalar models whose curved background extensions maintain second-order field equations and stress-tensors,''
Phys. Rev. D \textbf{80} (2009), 064015
\href{https://arxiv.org/pdf/0906.1967.pdf}{[arXiv:0906.1967 [gr-qc]]}.

\bibitem{Covariant_Galileon}
C.~Deffayet, G.~Esposito-Farese and A.~Vikman,
``Covariant Galileon,''
Phys. Rev. D \textbf{79} (2009), 084003
\href{https://arxiv.org/pdf/0901.1314.pdf}{[arXiv:0901.1314 [hep-th]]}.

\bibitem{Koba_G-infl}
T.~Kobayashi, M.~Yamaguchi and J.~Yokoyama,
``Generalized G-inflation: Inflation with the most general second-order field equations,''
Prog. Theor. Phys. \textbf{126}, 511-529 (2011)
\href{https://arxiv.org/pdf/1105.5723.pdf}{[arXiv:1105.5723 [hep-th]]}.

\bibitem{Koba_review}
T.~Kobayashi,
``Horndeski theory and beyond: a review,''
Rept. Prog. Phys. \textbf{82}, no.8, 086901 (2019)
\href{https://arxiv.org/pdf/1901.07183.pdf}{[arXiv:1901.07183 [gr-qc]]}.

\bibitem{NEC} V.~A.~Rubakov,
``The Null Energy Condition and its violation,''
Phys. Usp. \textbf{57} (2014), 128-142
\href{https://arxiv.org/pdf/1401.4024.pdf}{[arXiv:1401.4024 [hep-th]]}.

\bibitem{Creminelli:2006}
P.~Creminelli, M.~A.~Luty, A.~Nicolis and L.~Senatore,
``Starting the Universe: Stable Violation of the Null Energy Condition and Non-standard Cosmologies,''
JHEP \textbf{12} (2006), 080
\href{https://arxiv.org/pdf/hep-th/0606090.pdf}{[arXiv:hep-th/0606090 [hep-th]]}.

\bibitem{Creminelli:2010}
P.~Creminelli, A.~Nicolis and E.~Trincherini,
``Galilean Genesis: An Alternative to inflation,''
JCAP \textbf{11} (2010), 021
\href{https://arxiv.org/pdf/1007.0027.pdf}{[arXiv:1007.0027 [hep-th]]}.

\bibitem{Creminelli:2012}
P.~Creminelli, K.~Hinterbichler, J.~Khoury, A.~Nicolis and E.~Trincherini,
``Subluminal Galilean Genesis,''
JHEP \textbf{02} (2013), 006
\href{https://arxiv.org/pdf/1209.3768.pdf}{[arXiv:1209.3768 [hep-th]]}.

\bibitem{Cai:2012}
Y.~F.~Cai, D.~A.~Easson and R.~Brandenberger,
``Towards a Nonsingular Bouncing Cosmology,''
JCAP \textbf{08} (2012), 020
\href{https://arxiv.org/pdf/1206.2382.pdf}{[arXiv:1206.2382 [hep-th]]}.

\bibitem{DPSV}
C.~Deffayet, O.~Pujolas, I.~Sawicki and A.~Vikman,
``Imperfect Dark Energy from Kinetic Gravity Braiding,''
JCAP \textbf{10}, 026 (2010)
\href{https://arxiv.org/pdf/1008.0048.pdf}{[arXiv:1008.0048 [hep-th]]}.

\bibitem{G-bounce2011}
D.~A.~Easson, I.~Sawicki and A.~Vikman,
``G-Bounce,''
JCAP \textbf{11} (2011), 021
\href{https://arxiv.org/pdf/1109.1047.pdf}{[arXiv:1109.1047 [hep-th]]}.

\bibitem{Rubakov:2015}
V.~A.~Rubakov,
``Can Galileons support Lorentzian wormholes?,''
Teor. Mat. Fiz. \textbf{187}, no.2, 338-349 (2016)
\href{https://arxiv.org/pdf/1509.08808.pdf}{[arXiv:1509.08808 [hep-th]]}.

\bibitem{1601R} V.~A.~Rubakov,
``More about wormholes in generalized Galileon theories,''
Theor. Math. Phys. \textbf{188} (2016) no.2, 1253-1258
\href{https://arxiv.org/pdf/1601.06566.pdf}{[arXiv:1601.06566 [hep-th]]}.

\bibitem{Kobayashi:odd}
T.~Kobayashi, H.~Motohashi and T.~Suyama,
``Black hole perturbation in the most general scalar-tensor theory with second-order field equations I: the odd-parity sector,''
Phys. Rev. D \textbf{85} (2012), 084025
[erratum: Phys. Rev. D \textbf{96} (2017) no.10, 109903]
\href{https://arxiv.org/pdf/1202.4893.pdf}{[arXiv:1202.4893 [gr-qc]]}.

\bibitem{Kobayashi:even}
  T.~Kobayashi, H.~Motohashi and T.~Suyama,
  ``Black hole perturbation in the most general scalar-tensor theory with second-order field equations II: the even-parity sector,''
  Phys.\ Rev.\ D {\bf 89} (2014) no.8,  084042
 \href{https://arxiv.org/pdf/1402.6740.pdf}{[arXiv:1402.6740 [gr-qc]]}.

\bibitem{Arina_ani}
S.~A.~Mironov and A.~M.~Shtennikova,
``Perturbations in Horndeski Theory above Anisotropic Cosmological Background,''
JETP Lett. \textbf{119}, no.5, 339-344 (2024)
\href{https://arxiv.org/pdf/2305.19171.pdf}{[arXiv:2305.19171 [gr-qc]]}.

\bibitem{Olegi}
O.~A.~Evseev and O.~I.~Melichev,
``No static spherically symmetric wormholes in Horndeski theory,''
Phys. Rev. D \textbf{97}, no.12, 124040 (2018)
\href{https://arxiv.org/pdf/1711.04152.pdf}{[arXiv:1711.04152 [gr-qc]]}.

\bibitem{Cosmologies:2016}
M.~Libanov, S.~Mironov and V.~Rubakov,
``Generalized Galileons: instabilities of bouncing and Genesis cosmologies and modified Genesis,''
JCAP \textbf{08}, 037 (2016)
\href{https://arxiv.org/pdf/1605.05992.pdf}{[arXiv:1605.05992 [hep-th]]}.

\bibitem{Kobayashi_no-go}
T.~Kobayashi,
``Generic instabilities of nonsingular cosmologies in Horndeski theory: A no-go theorem,''
Phys. Rev. D \textbf{94}, no.4, 043511 (2016)
\href{https://arxiv.org/pdf/1606.05831.pdf}{[arXiv:1606.05831 [hep-th]]}.

\bibitem{math_nogo}
S.~Mironov,
``Mathematical Formulation of the No-Go Theorem in Horndeski Theory,''
Universe \textbf{5} (2019) no.2, 52

\bibitem{review_nogo}
S.~Mironov and V.~Volkova,
``Non-singular cosmological scenarios in scalar-tensor theories and their stability: a review,''
\href{https://arxiv.org/pdf/2409.16108.pdf}{[arXiv:2409.16108 [gr-qc]]}.

\bibitem{Kolevatov_2_scalar}
R.~Kolevatov and S.~Mironov,
``Cosmological bounces and Lorentzian wormholes in Galileon theories with an extra scalar field,''
Phys. Rev. D \textbf{94} (2016) no.12, 123516
\href{https://arxiv.org/pdf/1607.04099.pdf}{[arXiv:1607.04099 [hep-th]]}.

\bibitem{Koba_multi_gal}
S.~Akama and T.~Kobayashi,
``Generalized multi-Galileons, covariantized new terms, and the no-go theorem for nonsingular cosmologies,''
Phys. Rev. D \textbf{95} (2017) no.6, 064011
\href{https://arxiv.org/pdf/1701.02926.pdf}{[arXiv:1701.02926 [hep-th]]}.

\bibitem{Regge}
  T.~Regge and J.~A.~Wheeler,
  ``Stability of a Schwarzschild singularity,''
  Phys.\ Rev.\  {\bf 108} (1957) 1063.

\bibitem{odd_dynamic}
S.~Mironov, M.~Sharov and V.~Volkova,
``Linear stability of a time-dependent, spherically symmetric background in beyond Horndeski theory and the speed of gravity waves,''
\href{https://arxiv.org/pdf/2408.01480.pdf}{arXiv:2408.01480 [gr-qc]]}.

\bibitem{Kaluza_odd}
S.~Mironov, M.~Sharov and V.~Volkova,
``Time-dependent, spherically symmetric background in Kaluza-Klein compactified Horndeski theory and the speed of gravity waves,''
\href{https://arxiv.org/pdf/2408.06329.pdf}{[arXiv:2408.06329 [gr-qc]]}.

\bibitem{Takahashi:2016dnv}
K.~Takahashi and T.~Suyama,
``Linear perturbation analysis of hairy black holes in shift-symmetric Horndeski theories: Odd-parity perturbations,''
Phys. Rev. D \textbf{95} (2017) no.2, 024034
\href{https://arxiv.org/pdf/1610.00432.pdf}{ [arXiv:1610.00432[gr-qc]]}.

\bibitem{Babichev_sol}
E.~Babichev, C.~Charmousis and N.~Lecoeur,
``Exact black hole solutions in higher-order scalar-tensor theories,''
\href{https://arxiv.org/pdf/2309.12229.pdf}{ [arXiv:2309.12229 [gr-qc]]}.

\bibitem{Ogawa}
H.~Ogawa, T.~Kobayashi and T.~Suyama,
``Instability of hairy black holes in shift-symmetric Horndeski theories,''
Phys. Rev. D \textbf{93} (2016) no.6, 064078
\href{https://arxiv.org/pdf/1510.07400.pdf}{ [arXiv:1510.07400  [gr-qc]]}.

\bibitem{DPSV:1708}
R.~Kolevatov, S.~Mironov, V.~Rubakov, N.~Sukhov and V.~Volkova,
``Perturbations in generalized Galileon theories,''
Phys. Rev. D \textbf{96} (2017) no.12, 125012
\href{https://arxiv.org/pdf/1708.04262.pdf}{[arXiv:1708.04262 [hep-th]]}.

\bibitem{DPSV:1712}
S.~Mironov and V.~Volkova,
``Properties of perturbations in beyond Horndeski theories,''
Int. J. Mod. Phys. A \textbf{33} (2018) no.27, 1850155
\href{https://arxiv.org/pdf/arXiv:1712.09909.pdf}{[arXiv:1712.09909 [hep-th]]}.

\bibitem{DPSV_2023}
S.~Mironov and V.~Volkova,
``DPSV trick for spherically symmetric backgrounds,''
Nucl. Phys. B \textbf{1004}, 116550 (2024)
\href{https://arxiv.org/pdf/2306.17791.pdf}{[arXiv:2306.17791 [hep-th]]}.

\bibitem{ADM_Langlois}
D.~Langlois and K.~Noui,
``Hamiltonian analysis of higher derivative scalar-tensor theories,''
JCAP \textbf{07}, 016 (2016)
\href{https://arxiv.org/pdf/1512.06820.pdf}{[arXiv:1512.06820 [gr-qc]]}.

\bibitem{Langlois_review}
D.~Langlois,
``Dark energy and modified gravity in degenerate higher-order scalar\textendash{}tensor (DHOST) theories: A review,''
Int. J. Mod. Phys. D \textbf{28} (2019) no.05, 1942006
\href{https://arxiv.org/pdf/1811.06271.pdf}{[arXiv:1811.06271 [gr-qc]]}.

\bibitem{Cuscuton}
N.~Afshordi, D.~J.~H.~Chung and G.~Geshnizjani,
``Cuscuton: A Causal Field Theory with an Infinite Speed of Sound,''
Phys. Rev. D \textbf{75} (2007), 083513
\href{https://arxiv.org/pdf/0609150.pdf}{[arXiv:hep-th/0609150 [hep-th]]}.

\bibitem{1812wormhole}
S.~Mironov, V.~Rubakov and V.~Volkova,
``More about stable wormholes in beyond Horndeski theory,''
Class. Quant. Grav. \textbf{36} (2019) no.13, 135008
\href{https://arxiv.org/pdf/1812.07022.pdf}{[arXiv:1812.07022 [hep-th]]}.

\bibitem{2212wormhole}
S.~Mironov, V.~Rubakov and V.~Volkova,
``In hot pursuit of a stable wormhole in beyond Horndeski theory,''
Phys. Rev. D \textbf{107} (2023) no.10, 104061
\href{https://arxiv.org/pdf/2212.05969.pdf}{[arXiv:2212.05969 [gr-qc]]}.

\bibitem{BH_Langlois2014}
J.~Gleyzes, D.~Langlois, F.~Piazza and F.~Vernizzi,
``Healthy theories beyond Horndeski,''
Phys. Rev. Lett. \textbf{114} (2015) no.21, 211101
\href{https://arxiv.org/pdf/1404.6495.pdf}{[arXiv:1404.6495 [hep-th]]}.

\bibitem{cadabra} 
K. Peeters, Cadabra2: computer algebra for field theory revisited, J. Open Source Softw. 3 (2018), no. 32 1118.

\end{thebibliography}
\end{document}